\documentclass[letterpaper, 10 pt, conference]{ieeeconf}

\IEEEoverridecommandlockouts
\usepackage{amsthm}
\usepackage{cite}
\usepackage{microtype}
\usepackage{graphicx}
\usepackage{subfigure}
\usepackage{booktabs} 
\usepackage{algorithm}
\usepackage{algorithmic}
\usepackage[hidelinks]{hyperref}
\usepackage{url}
\usepackage{dsfont} 
\usepackage{siunitx}
\usepackage{amsmath,amssymb,amsfonts}
\usepackage{bm}      
\usepackage{cleveref}
\usepackage{mathrsfs}
\usepackage{hyperref}
\usepackage{multirow}
\usepackage{amsthm}
\usepackage{qrcode}
\theoremstyle{plain}
\usepackage{graphicx}

\newtheorem{theorem}{Theorem}

\title{\LARGE \bf Merging Parameter Estimation and Classification Using LASSO
}

\author{Le Wang, Ying Wang, Yu Qiu, Mian Li, and Håkan Hjalmarsson
\thanks{*This work was supported by VINNOVA Competence
Center AdBIOPRO, contract [2016-05181], by the
Swedish Research Council through the research environment
NewLEADS (New Directions in Learning Dynamical Systems),
contract [2016-06079], and contract 2019-04956. This work was also supported by the National Natural Science Foundation of China under
Grant 52275263.}
\thanks{Le Wang, Ying Wang, and Håkan Hjalmarsson are with Division of Decision and Control Systems, School of Electrical Engineering and Computer Science, KTH Royal Institute of Technology, 10044 Stockholm, Sweden. Le Wang is also with UM-SJTU Joint Institute, Shanghai Jiao Tong University, 200240 Shanghai, China. Håkan Hjalmarsson is also with Centre for Advanced Bio Production, 10044 Stockholm, Sweden. Email: \{le6, yinwang, hjalmars\}@kth.se.}
\thanks{Yu Qiu is with SAIC Motor RD Innovation Headquarters, 201807
Shanghai, China. Email: qiuyu\_1123@163.com.}
\thanks{Mian Li is with Global Institute of Future Technology, Shanghai Jiao Tong University, 200240 Shanghai, China. Email: mianli@sjtu.edu.cn.}
}

\begin{document}

\maketitle
\thispagestyle{empty}
\pagestyle{empty}

\begin{abstract}
Soft sensing is a way to indirectly obtain information of signals for which direct sensing is difficult or prohibitively expensive. It may not \textit{a priori} be evident which sensors provide useful information about the target signal, and various operating conditions often necessitate different models. In this paper, we provide a systematic method to construct a soft sensor that can deal with these issues. We propose a single estimation criterion, where the objectives are encoded in terms of model fit, model sparsity (reducing the number of different models), and
model parameter coefficient sparsity (to exclude irrelevant sensors). 
The proposed method is tested on real-world scenarios involving prototype vehicles, demonstrating its effectiveness.
\end{abstract}

\section{Introduction}

Estimation of wheel forces in vehicles is an important problem \cite{risaliti2016virtual}. For instance, in the optimization of vehicle design, understanding the forces acting on each wheel and the load-bearing conditions allows for the enhancement of the vehicle's suspension system, braking system, and overall operational performance. Moreover, it contributes to vehicle safety assessments and vehicle fault diagnostics, resulting in the extension of vehicle lifespan and the reduction of the risk of malfunctions \cite{zhu2019takagi}. Nevertheless, wheel force transducers are both costly and challenging to be installed on vehicles operating on public roads. Consequently, alternative cost-effective methods for accurate torque estimation, such as soft sensor approaches, have surfaced and garnered significant attention.

Soft sensor technology involves estimating difficult-to-measure variables by utilizing more easily measurable variables as inputs to specially designed mathematical models \cite{jiang2020review,wang2023edge,wang2019soft, fang20203,wang2023sensor}. There are different categories of soft sensors, such as data-driven and model-driven soft sensors \cite{kadlec2009data,kang2017fast}. Here we consider a data-driven technique. The first step is to use historical data to identify a model of the relationship between the measured variables and the desired (unmeasured) variable \cite{jiang2020review}. We use multi-input single-output (MISO) finite impulse response (FIR) models because this aligns with our target application. Moreover, these models are simple to use as they can be estimated using least-squares techniques \cite{Ljung:99}. One specific MISO-FIR model can be identified based on a dataset corresponding to a specific operating condition. However, in this paper we consider vehicles running on various road conditions \cite{wang2023hybrid}, and therefore multiple models may be required to cover the envelope of all working conditions. This can increase the computational load on the edge devices (in which the soft sensor models are implemented) installed on the vehicles. Thus, it prompts the need for merging different, but similar, working conditions into one model. In practical applications, there are often sensors with little or no relevance to the quantity of interest, making it challenging to directly classify models. Thus, there is a need for methods able to cope with such settings.

In the literature, many researchers considered various settings. For example, \cite{zheng2023data} proposed a data-driven method to select the K most relevant sensors among S candidate ones that best fit the response of one vertical wheel force; \cite{rao2015classification} proposed sparse group lasso to do a less restrictive form of structured sparse feature selection; \cite{ghosh2005classification} proposed a LASSO-based variable selection algorithm based on linear combinations of the gene expression. In \cite{huo2020sgl}, a tumor classification method based on sparse group lasso and support vector machine was introduced. However, reducing the number of different models and excluding irrelevant sensors are not taken into account at the same time in those works.

This paper proposes an FIR-based modified fused lasso algorithm to classify multiple working conditions with irrelevant sensors into a set of models, excluding irrelevant sensors and merging together similar working conditions. The presence of irrelevant sensors is handled by employing a LASSO-type of regularization measuring the distance between models. To reduce the number of different models, we use the parameter difference $\ell_2$-norm method. Then, we use unsupervised clustering methods to cluster similar working conditions.

\begin{figure*}[ht]
    \centering
    \includegraphics[width=1.95\columnwidth]{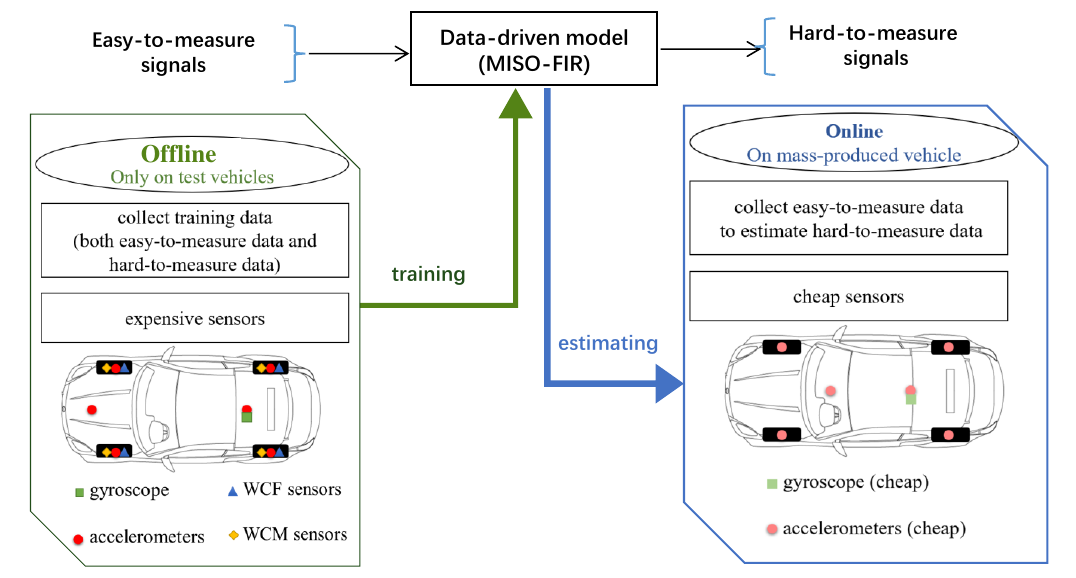}
    \caption{\centering
    Working principle for online stage and offline stage.}
    \label{fig:ps2}
\end{figure*}

The contributions of this work are summarized as follows:

\begin{itemize}
  \item A parameter difference $\ell_2$-norm $\&$ LASSO regularization algorithm based on a MISO-FIR model and K-means is proposed to determine the irrelevant sensors and optimize the model parameters. To help with the tuning of the weighting factor of the penalty term we provide an upper bound.
  \item The proposed method is tested in real-world scenarios with data obtained from prototype vehicles. The results support the applicability of the method to real data applications, and indicate that the method may be of value in research on vehicle production.
\end{itemize}

To the best of our knowledge, it is the first attempt to apply this kind of algorithm to the vehicle working condition classification problem with irrelevant sensors.

The rest of this paper is organized as follows. Section~II introduces
problem statement. Section III presents the methodology. Sections IV shows the experiments setup and results. Finally, Section V concludes this paper.

\vspace{2pt}

\section{Problem Statement}\label{sec:problem}

This paper focuses on how to estimate ``lean" soft sensors, valid for a wide range of working conditions. By ``lean", we mean that the soft sensor uses no more sensors (inputs) than necessary for any given working condition and, further, when possible the same configuration of sensors is used for different working conditions.

The principle and the background of this paper are presented in Fig. \ref{fig:ps2}. We use easy-to-measure sensor data (i.e., accelerometers) and hard-to-measure sensor data (i.e., wheel center force data, obtained from WCF sensors) to train the data-driven models offline. For the online stage, we just apply easy-to-measure sensor data to the model to estimate the hard-to-measure quantities.

Multiple-input single-output finite impulse response (MISO-FIR) models are used, because this aligns with our target application. The output of such a model can be expressed as a linear combination of a specific set of input variables \cite{Ljung:99} as follows,
\begin{equation}
\footnotesize
     y(t)=\mathbf{a}_{0}^\mathrm{T} \mathbf{x}(t)+\dots+\mathbf{a}_{n}^\mathrm{T} \mathbf{x}(t-n)+\boldsymbol{\varepsilon}(t), \quad t=n+1, \dots, M+n,
\label{equ:linear}
\end{equation}
where $n$ is the order of the model, $M$ is the length of ${y}(t)$, $\mathbf{x}(t) \in \mathbb{R}^m$ and ${y}(t) \in \mathbb{R}$ respectively are inputs and output of the model, $\mathbf{a}_{i} \in \mathbb{R}^m$, $i=0, \dots, n$, are unknown parameters to be estimated, and ${\boldsymbol{\varepsilon}(t)}$ is the measurement noise of the model, which we assume is independent and identically distributed.

We can re-write \eqref{equ:linear} compactly as

\begin{equation}
    \mathbf{Y}=\bm{\Phi} \bm{\theta}+\mathbf{E},
    \label{equ:vector1}
\end{equation}
where the model parameter
$\bm{\theta}=\left[\mathbf{a}_{0}^\mathrm{T}, \dots,\mathbf{a}_{n}^\mathrm{T}\right]^\mathrm{T}$, and
\begin{gather*}
\scriptsize
    \mathbf{Y}=\left[\begin{array}{c}
    y(n+1) \\
    \vdots \\
    y(M+n)
    \end{array}\right],
    \bm{\Phi}=\left[\begin{array}{c}
    \phi^\mathrm{T}(n+1) \\
    \vdots \\
    \phi^\mathrm{T}(M+n)
    \end{array}\right],
    \mathbf{E}=\left[\begin{array}{c}
    \varepsilon(n+1) \\
    \vdots \\
    \varepsilon(M+n)
    \end{array}\right]
\end{gather*}
with the regressor $\phi(t)=\left[\mathbf{x}^\mathrm{T}(t), \dots, \mathbf{x}^\mathrm{T}(t-n)\right]^\mathrm{T}$.

\vspace{3pt}

For each working condition, we may construct one data-driven model. However, when there are hundreds of working conditions, we need to build hundreds of data-driven models. This will result in a heavy computational burden on edge devices (in which the soft sensor models have been implemented). Thus, we would like to reduce the computational burden and still keep the estimation accuracy high. This requires us to reduce the number of different models. Meanwhile, we need to consider the model parameter coefficient sparsity since irrelevant sensors exist. Furthermore, we do not know which working conditions that can be described by the same model. Thus, the difficulties of this paper lie in:

\begin{itemize}
  \item It is an unsupervised learning problem because there are no labels for categories.
  \item It is a big data problem, subject to dynamics and high-dimension.
\end{itemize}

\vspace{1pt}

\section{Methodology}\label{sec:method}
In this section, we first introduce the proposed method. Then, we discuss the choice of hyperparameters. Finally, the clustering part is discussed followed by a summary of the algorithm.

\subsection{Parameter Difference $\ell_2$-norm $\&$ LASSO Regularization}

As mentioned in Section \ref{sec:problem}, the challenge lies in having multiple working conditions, where some of the conditions may be modelled using the same model, but where we do not know which conditions belong together. For each working condition, indexed by $k=1,\dots,K$, for which a data set $Y_k,\Phi_k$ has been collected, one can estimate a FIR model using least-squares as

\begin{equation}
    \hat{\theta}_k = \mathop{\rm{arg} \min}_{\theta_k}  ||Y_k-\Phi_k \theta_{k}||_2^2, \quad k=1,\dots,K.
     \label{equ:least}
\end{equation}

In practical applications, such as the one considered in this paper, the number of working conditions K can be very large. In our case it is in the order of hundreds. One task is now to eliminate sensors that do not significantly contribute to a model. This corresponds to determining which elements of $\theta_k$ are zero. Another task is to determine for which working conditions the same model can be used. As already mentioned, this is important in order to reduce the computational load of the soft sensor itself. Instead of performing the three tasks of model parameter estimation, sensor elimination and model clustering separately, we employ one single  criterion where each of these requirements is encoded in a separate term: i)  a term measuring the total squared model error, ii) a term measuring the distance between the different models, and iii) a term measuring the importance of each sensor. For the third term, we use the $\ell_1$-norm, i.e., LASSO, as it is known to lead to sparse solutions \cite{Tibshirani:96}. For the second term, we use the Euclidean distance between the parameter vectors of different models. In comparison with group lasso \cite{Yuan:06}, rather than forcing an entire parameter vector to be zero, we are in this way forcing the difference between parameter vectors to be zero, thus achieving that the models are clustered.
In summary, we propose the following criterion

\begin{equation}
\footnotesize
    \mathop{\min}_{\theta_{1} \dots\theta_{K}} \sum_{k=1}^K \|Y_k-\Phi_k \theta_{k}\|_2^2 +\frac{1}{2} \lambda_1\sum_{k,i=1}^{K,K}\|\theta_{k}-\theta_{i}\|_2 +\lambda_2 \sum_{k,j=1}^{K,J}\|\theta_{k}^j\|_1,
     \label{equ:obj}
\end{equation}
where, $\lambda_1$, and $\lambda_2$ are positive constants that are used to control the trade-off among the model fit (the first term), the clustering (the second term), and sparsity (the third term) of the model parameters. $\lambda_1$ is the penalty term coefficient of the inter-group which reduces the number of different models, while $\lambda_2$ is the penalty term coefficient of the intra-group (which shows the sparsity of model parameters within one model). Here, $J$ is the type of different sensors inside one working condition. For example, $\theta_k^j$ is the parameter of working condition $k$ and sensor channel $j$. $Y_k$ is the value of the system output, $\Phi_k$ is the regressor of the model, and $\theta_k=[\theta_k^1,\dots,\theta_k^j]^{\rm{T}}$.


\subsection{Choosing the Hyperparameters }

Recall that the objective is to eliminate sensors that do not contribute significantly to the estimate of the target signal, and to group together working conditions that can be modeled in the same way. To this end, the parameter $\lambda_2$ controls the sparsity and the parameter $\lambda_1$ how many different parameter vectors are forced to resemble one another.  Picking these parameters too low does not achieve these objectives, while too large values sacrifices the model fit as given by the first term in (4). It is thus necessary to find a good balance between these conflicting objectives. Starting with $\lambda_1$, we have the following result.

\begin{theorem}

Let $Y_k \in \mathbb{R}^{M_k}$, $\Phi_k \in \mathbb{R}^{M_k \times n_{\theta}}$, $\theta_k \in \mathbb{R}^{n_{\theta}}, k=1,\dots,K$, where $n_{\theta}=n \times J$, $M_k$ is the size of $Y_k$.
Define

\begin{equation}
\lambda_{1\rm{max}}=\max_{k\in \{1,\dots,K\}}\left\{\frac{2} {K-1} \| \Phi_k^{\rm{T}}[Y_k-\Phi_k \theta_*]\|_{2} \right\},
\end{equation}
where
\begin{equation}
    \theta_{\star}=\mathop{\rm{arg}\min}_{\theta} \sum_{k=1}^{K} \|Y_k-\Phi_k\theta\|_2^2.
    \end{equation}

Given that $\lambda_1 \geq \lambda_{1\rm{max}}$ and $\Phi_k^ {\rm{T}} \Phi_k \succ 0$, $\forall k \in {1,\dots, K}$, it holds that the solution is given by

\begin{equation}
    \theta_k^*=\theta_{\star}, k=1,\dots,K.
    \end{equation}

\end{theorem}


Theorem 1 gives an upper bound for $\lambda_1$ that can be easily computed. There is a corresponding upper bound for $\lambda_2$, given by $\lambda_{2\rm{max}}= \max \left\{2||\Phi_k^{\rm{T}} Y_k||_\infty \right\}, k=1,\dots,K $. For $\lambda_2>\lambda_{2\rm{max}}$, it holds that $\theta_k^*=0$ for all $k$. In other words, in order to get the nontrivial $\theta_k^*\neq0$, we need to choose $\lambda_2\leq\lambda_{2\rm{max}}$.

To find appropriate penalty coefficients, we use grid search. To select these hyperparameters, we estimate the model parameters $\theta_k$ on an estimation dataset, and then pick $\lambda_1$ and $\lambda _2$ that minimize (4) on a validation dataset. We use logaritmically spaced grids $\lambda_1=[0.00001\lambda_{1\rm{max}},\dots, 0.1\lambda_{1\rm{max}},\lambda_{1\rm{max}}]$, $\lambda_2=[0,10^{-6},10^{-4},10^{-2},1,10^2]$ (which are inside the boundary of $\lambda_2$).

\subsection{Clustering}

While the proposed algorithm forces models corresponding to different operating points to become close (due to the term for which $\lambda_1$ is a factor), the parameters may not become identical since the same tuning parameter $\lambda_1$ is used for all parameter differences. Therefore, we use a post-processing step where a clustering algorithm is used to merge models. While there is a wide range of clustering algorithms available, we use K-means in the experimental section. A potential improvement of the algorithm would be to tune the weights for different parameter differences $\|\theta_k-\theta_i\|_2$ individually. However, this would dramatically increase the number of hyperparameters that require tuning.

\subsection{Summary}

\begin{itemize}
  \item Step 1: Calculate the boundary of the inter-group penalty term coefficient $\lambda_{1\rm{max}}$ and the intra-group penalty term coefficient $\lambda_{2\rm{max}}$.
  \item Step 2: Use grid search to choose the two hyperparameters $\lambda_1$ and $\lambda_2$.
  \item Step 3: Estimate the model parameter $\theta_k$ using hyperparameters obtained from the step before.
  \item Step 4: Use k-means to cluster different kinds of models.
\end{itemize}

\section{Experiments}\label{sec:simulation}
\subsection{Experimetal Setup}

\begin{figure}[htbp!]
    \centering
\includegraphics[width=0.7\linewidth]{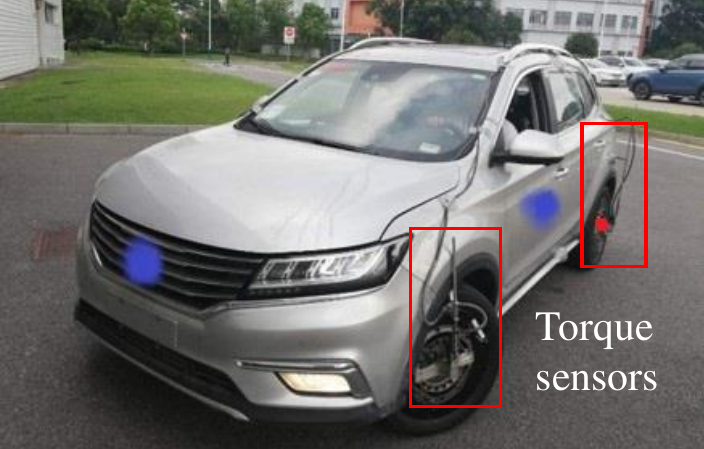}
    \caption{The torque sensors installed on the field test vehicle.}
    \label{fig:car}
\end{figure}

\begin{figure}[htbp!]
    \centering
\includegraphics[width=0.7\linewidth]{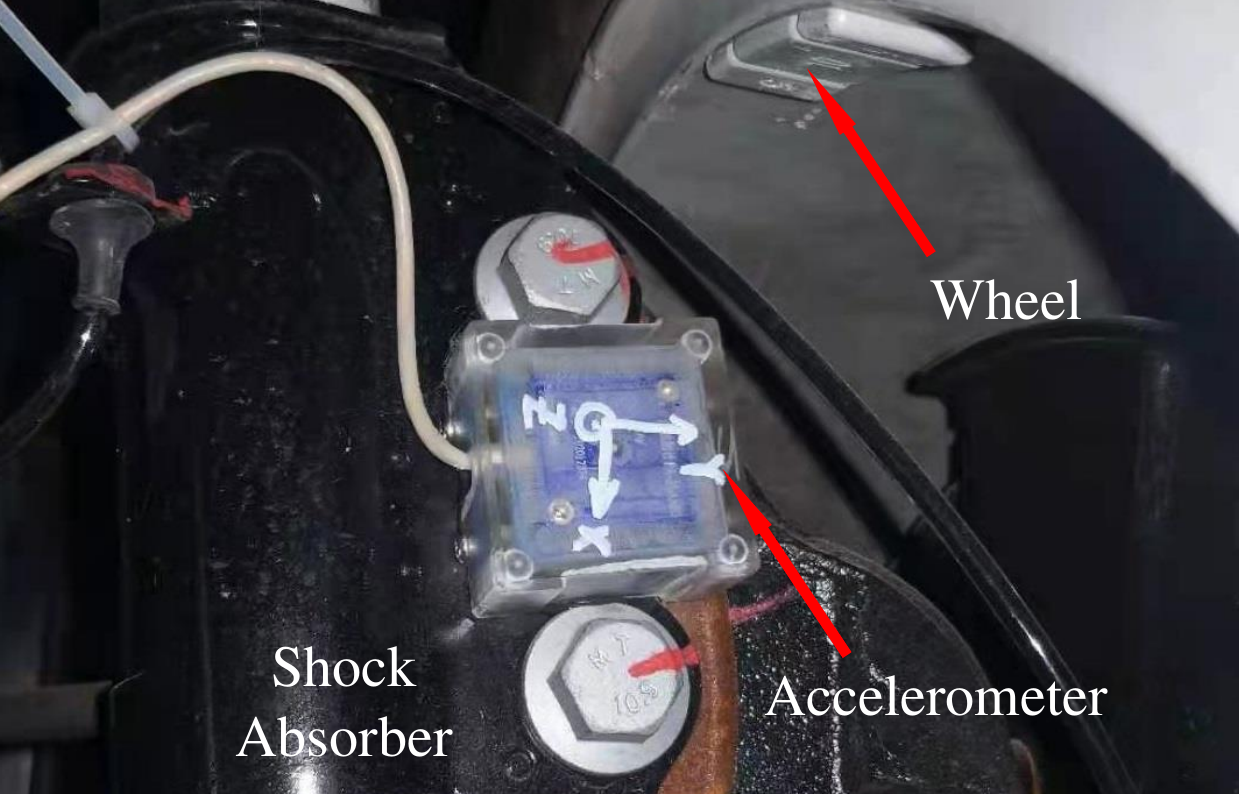}
    \caption{The accelerometers (three-axis) installed on wheels.}
    \label{fig:xyz}
\end{figure}

All the experiments are conducted using a 1.60 GHz Intel Core i5-8250U processor. The algorithm implementation is done in Matlab 2021b. The data used in this work is collected from multiple accelerometers and torque sensors installed on a prototype PT-4WD vehicle. In Fig. \ref{fig:car}, it shows the vehicle used for field tests. The red boxes indicate the places where the torque sensors are installed, which can measure the torque in X, Y, and Z directions when driving. In Fig. \ref{fig:xyz}, it shows the 3-axis accelerometers installed on wheels.

The datasets used in this work are named, for example,  ``BR30-1", where the ``BR" refers to the braking working condition, ``30" is the speed of the vehicle (i.e., 30km/h), and ``1" is the index of the datasets (i.e., the first set under this condition). All datasets used in this work follow this format and each set includes system signals obtained from the multiple sensors during the field tests. In this paper, we use 3 working conditions randomly chosen from the hundreds of available datasets to demonstrate the effectiveness of the proposed algorithm, namely: BR30, BR40, and WBA40, where WBA is the abbreviation of the ``Washboard Asynchronous" working condition.

For each working condition, we have hundreds of sensor categories. In this paper, 4 easy-to-measure sensors are chosen as the inputs to estimate the one hard-to-measure sensor. The input sensors and output sensor are shown in Table \ref{tab:sensor}. The model order is taken as n=50, which gives 200 parameters to estimate for each working condition as we have 4 inputs. The first 50 parameters are the parameters of the first sensor, whose number is the ``throttle pedal (number 57 sensor, see Tab.\ref{tab:sensor})". The parameters 51-100 represent the second sensor, which is the ``road surface roughness". Parameters 101-150 and 151-200 are ``Right front suspension displacement" and ``Left front suspension displacement", respectively.

\begin{table}[h]
  \centering
  \footnotesize
  \caption{Definitions of the input sensors and output sensor.}
  \begin{tabular}{ccc} 
    \toprule 
    &Sensor  & Category \\
& Number  &  \\

    \midrule 
    &57 & Throttle pedal\\
    Input&90  & Road surface roughness \\
    Sensors&91  & Right front suspension displacement\\
    &92  & Left front suspension displacement\\
    \midrule 
    Output&35  & Left front wheel center force in the Z direction\\
    \bottomrule 
  \end{tabular}
  \label{tab:sensor}
\end{table}

\subsection{Experimental Results}

\subsubsection{Parameter Comparison}

\begin{figure}[htbp!]
    \centering
\includegraphics[width=0.8\linewidth,height=5.3cm]{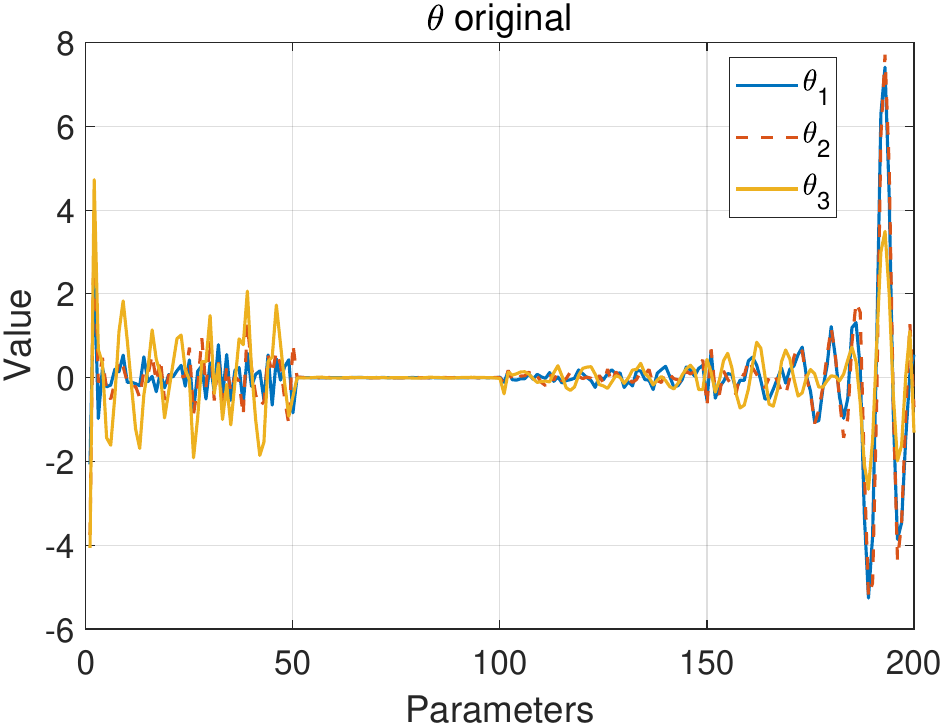}
\vspace{-0.2cm}
    \caption{Least-squares estimate of $\theta$ for the three used working conditions.}
    \label{fig:2}
\end{figure}

\begin{figure}[htbp!]
    \centering
\includegraphics[width=0.8\linewidth,height=5.3cm]{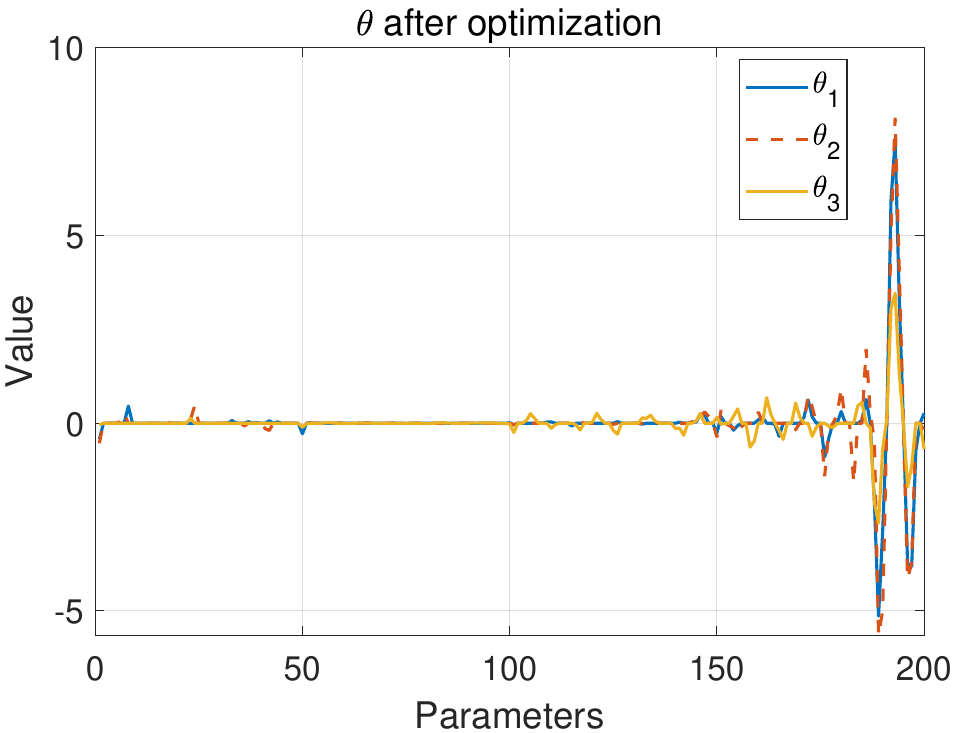}

    \caption{The resulting parameter estimates for the three used working conditions when the proposed algorithm is used.}
    \label{fig:3}
\end{figure}

Figure \ref{fig:2} shows the estimated parameter vector for the three different operating conditions when each parameter vector has been estimated separately using least-squares (3). The corresponding estimates when our proposed approach has been used are shown in Fig. \ref{fig:3}.

By comparing the two figures, it can be observed that the parameters of the first sensor have non-negligible values when standard least-squares estimation is used. However, they are nearly compressed to zero after applying our proposed algorithm. This indicates that the signal from this sensor contributes little to the estimation of the output sensor signal, and can therefore be considered an ``irrelevant sensor". The parameter values of the second sensor are nearly zero both before and after optimization, indicating that this sensor makes a minimal contribution to estimating the output sensor signal, which is also an ``irrelevant sensor". The third sensor has small values both before and after optimization, indicating a modest contribution to estimating the target sensor signal. The fourth sensor has large values both before and after optimization, making a significant contribution to estimating the target sensor signal. The penalty coefficients $\lambda_1$ and $\lambda_2$ are tuned as described in Section II, resulting in the inter-group penalty term coefficient $\lambda_1=0.00001\lambda_{1\rm{max}}=0.0369$, and the intra-group penalty term coefficient $\lambda_2=1$.


\subsubsection{Model Classification}

Different kinds of classification algorithms can be used to classify the models. Here, we use K-means, which is a basic clustering algorithm. We choose the cluster number $K=2$. Table \ref{tab} shows the classification results of the three working conditions. It shows that BR30 and BR40 are described by the same model, while WBA40 is classified to require another model.

\begin{table}[h]
  \centering
  \caption{Classification results of the three working conditions.}
  \begin{tabular}{cc} 
    \toprule 
    Working condition & Category \\
    \midrule 
    BR30 & Category 1\\
    BR40  & Category 1 \\
    WBA40  & Category 2\\
    \bottomrule 
  \end{tabular}
  \label{tab}
\end{table}

\subsubsection{Model Evaluation}

To evaluate the performance of the proposed approach, we use the goodness of FIT \cite{golub2013matrix}, defined as

\begin{equation}
    FIT=\left(1-\frac{\|Y-\hat{Y}\|_2^2}{\|Y-\bar{Y}\|_2^2}\right)\times 100\%,
    \label{equ:fit}
\end{equation}
where $Y$ are measurements collected from the sensors, $\hat{Y}$ is the estimated value obtained from the proposed algorithm, and $\bar{Y}$ is the mean vector of $Y$. The closer the value of FIT to $100\%$, the better the performance is.

To study how well the proposed method is able to classify the different working conditions (cf. Table II), we estimate one model each, using least-squares, for the three working conditions and then computed the FITs when each of these models is applied to validation data sets coming from all three different working conditions. The results of this can be found in the first three columns of Table III. From the first two block rows, it can be seen that working conditions BR30 and BR40 result in models that can be used for both BR30 and BR40, but not for WBA40. The last block row shows that the model estimated using WBA40-1 works well on validation data from the same working condition but not for BR30 and BR40. These results suggest that BR30 and BR40 can be represented with one common model, while a separate model is needed for WBA40. This is consistent with the outcome of the proposed method as indicated in Table \ref{tab}. Table \ref{tab:class} shows the effectiveness after classification, which verifies the classification results in Table \ref{tab}.


\begin{table}[h]
  \centering
  \footnotesize
  \caption{FIT on evaluation data}
  \begin{tabular}{cccc} 
    \toprule 
    Dataset for  & Dataset for which &  FIT using  & FIT using  \\
     estimated  &  the FIT  & least-square  & the proposed \\
     model & is computed & estimate (3) & method\\
    \midrule 
          & BR30-2 &76.49$\%$  & 76.44$\%$\\
       BR30-1&  BR40-2 & 75.46$\%$ &  74.19$\%$ \\
        &   WBA40-2 & -7.53$\%$&  44.12$\%$\\
         \midrule 
          & BR30-2 & 74.21$\%$ & 75.38$\%$\\
        BR40-1 & BR40-2 & 75.65$\%$ &  75.63$\%$\\

        & WBA40-2 & 5.80$\%$ & 46.61$\%$\\
         \midrule 
          & BR30-2 & 44.10$\%$  & -6.41$\%$\\
       WBA40-1 & BR40-2 & 46.58$\%$ & 6.33$\%$\\
         & WBA40-2 & 73.18$\%$ & 73.30$\%$ \\

    \bottomrule 
  \end{tabular}
  \label{tab:fit}
\end{table}

\begin{table}[h]
    \centering
     \footnotesize
    \caption{FIT on evaluation data after classification}
    \begin{tabular}{ccc}
    \toprule 
    Dataset for & Dataset for which the  & FIT using least-square \\
     estimated mode & FIT is computed & estimate (3)\\
    \midrule 
    \multirow{2}{*}{\centering BR30-01 + BR40-01} & BR30-02 & 75.38$\%$ \\
       &  BR40-02 & 75.52$\%$\\
        \midrule 
      WBA40-01 & WBA40-02 & 73.18$\%$\\
    \bottomrule 
    \end{tabular}

    \label{tab:class}
\end{table}

\begin{figure*}[h]

  \begin{minipage}{0.3\textwidth}
    \centering
    \includegraphics[scale=0.4,height=4cm]{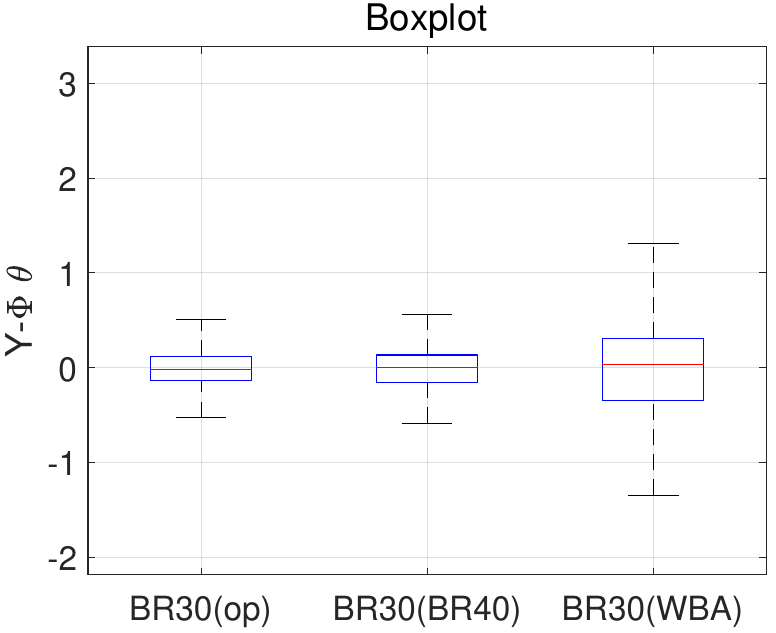}
    \caption{Boxplot for the BR30-2 working condition.}
    \label{fig:BR30box}
  \end{minipage}
  \hfill
  \begin{minipage}{0.3\textwidth}
    \centering
    \includegraphics[scale=0.4,height=4cm]{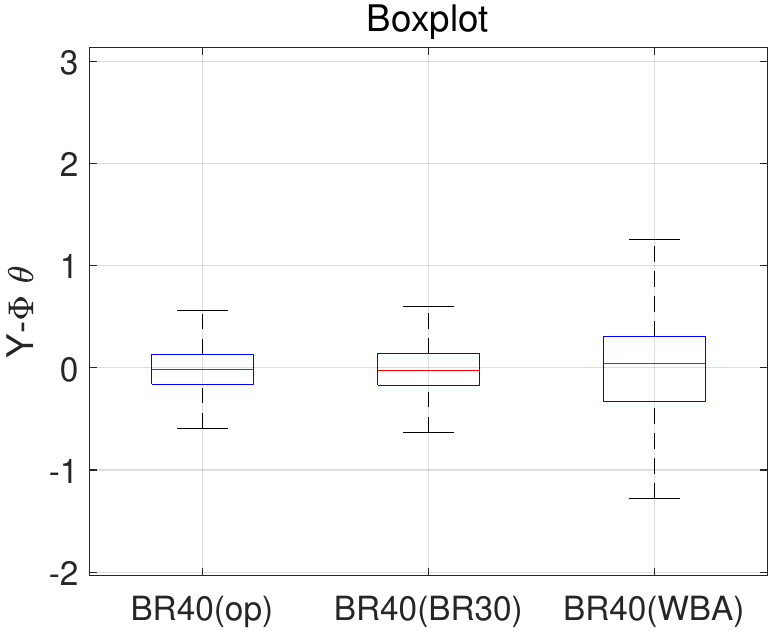}
    \caption{Boxplot for the BR40-2 working condition.}
    \label{fig:BR40box}
  \end{minipage}
  \hfill
  \begin{minipage}{0.3\textwidth}
    \centering
    \includegraphics[scale=0.4,height=4cm]{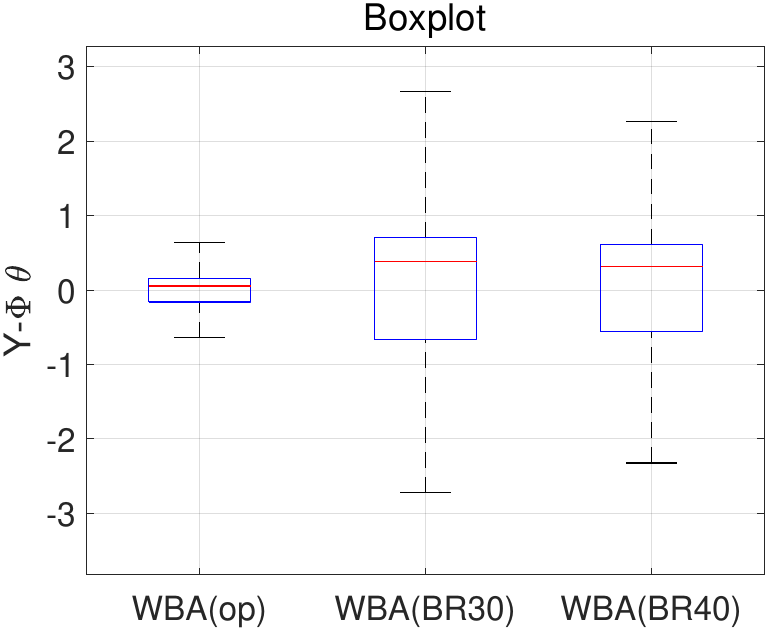}
    \caption{Boxplot for the WBA40-2 working condition.}
    \label{fig:WBAbox}
  \end{minipage}

  \medskip

\begin{minipage}{0.3\textwidth}
    \centering
    \includegraphics[scale=0.4]{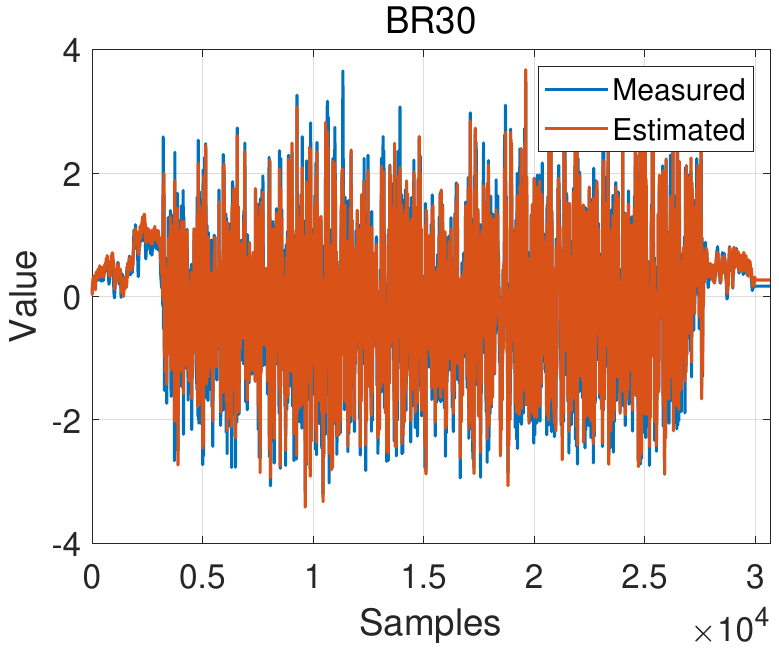}
    \caption{Comparison between the output signal (see Table 1) for the BR30-2 working condition, and the estimated output using the model obtained from the BR30-1 dataset. }
    \label{fig:BR30}
  \end{minipage}
  \hfill
  \begin{minipage}{0.3\textwidth}
    \centering
    \includegraphics[scale=0.4]{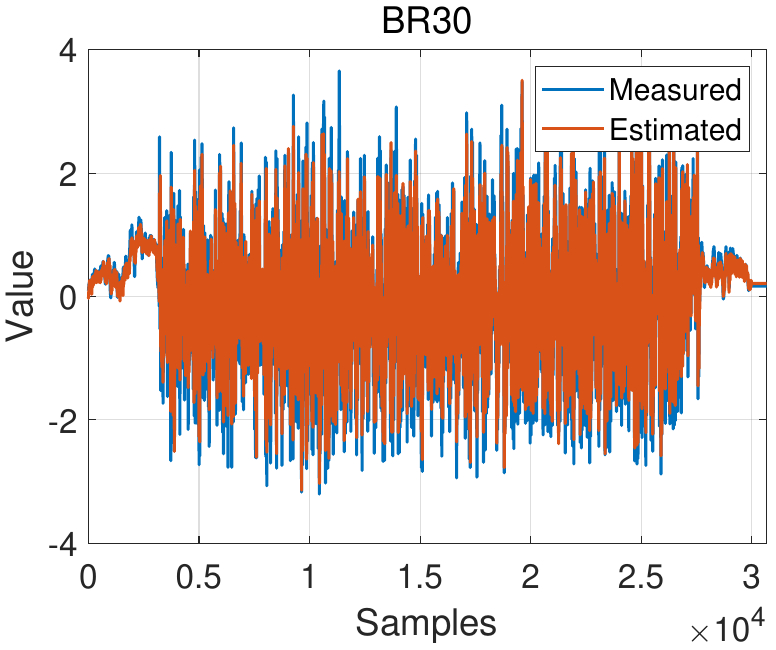}
    \caption{Comparison between the output signal (see Table 1) for the BR30-2 working condition, and the estimated output using the model obtained from the BR40-1 dataset. }
    \label{fig:30-40}
  \end{minipage}
  \hfill
  \begin{minipage}{0.3\textwidth}
    \centering
    \includegraphics[scale=0.4]{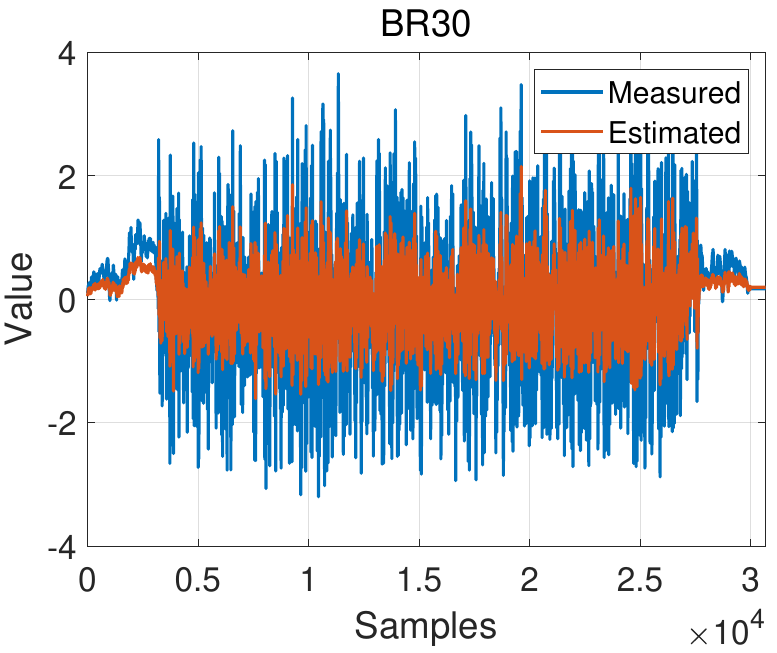}
    \caption{Comparison between the output signal (see Table 1) for the BR30-2 working condition, and the estimated output using the model obtained from the WBA40-1 dataset. }
    \label{fig:30-WBA}
  \end{minipage}

\end{figure*}

From Figs. \ref{fig:BR30box}--\ref{fig:WBAbox}, we can see the boxplot of the cross-evaluation performance of BR30-2, BR40-2, and WBA40-2 working conditions, respectively. The content inside the parentheses is the cross-evaluation target, for example, BR30 (BR40) means that use parameters of BR40 to estimate BR30. It can be observed that the performance of the cross-evaluation between BR30 and WBA40, BR40 and WBA40 is not satisfactory. In Figs. \ref{fig:BR30}--\ref{fig:30-WBA}, we can see the cross-evaluation between different working conditions. These figures show that the cross-evaluation effect between BR30 and BR40 is good, while the performance is poor between BR30 and WBA40. These observations support the results of the proposed method.

Alternatively to using the Euclidean norm for terms in the second sum in (4), one can use the squared Euclidean distance. This has been tested with similar results. Results show that $\ell_2$-norm with square just accelerates the algorithm within a small range. In this paper, we just use the $\ell_2$-norm for the inter-group.

\section{Conclusion}\label{sec:conclusion}
This article contributes with a LASSO-type algorithm based on MISO-FIR models within the soft sensor framework, to classify data from multiple working conditions into a smaller number of models, excluding irrelevant sensors from each model. Validation with field test datasets from a prototype vehicle shows good performance for the classification. The problem discussed in this paper is very common in practical applications such as the estimation of vehicle dynamics. For future work, we will extend the classification of working conditions to include more scenarios.


\bibliography{ref}
\bibliographystyle{IEEEtran}

\end{document}